# On the grain growth behavior of Ni-W alloys in the extremely fine-grained regime

Keerti Pandey[a,1], Aman Prasad[a,2]

[a]Department of Materials Engineering, Indian Institute of Science, Bangalore, Karnataka, India – 560012

[1]Currently at:

Department of Metallurgical Engineering & Materials Science, Indian Institute of Technology Bombay, Powai, Maharashtra, India – 400076
keertipandey30@gmail.com

[2]Currently at:

Institut des Matériaux de Nantes Jean Rouxel (IMN), Nantes Université, CNRS UMR 6502, Polytech Nantes, Rue Christian Pauc, BP 50609, CEDEX 3, 44306, Nantes, France.
amanprasad14@gmail.com

Address all correspondence to this author: e-mail keertipandey30@gmail.com,

## Abstract

Nanocrystalline pure FCC metals and some alloys are known to exhibit abnormal grain growth. Addition of solutes, such as W, has led to improved grain size stability in nanocrystalline Ni. While several groups have investigated grain growth behavior in Ni-based binary alloys, grain sizes greater than 15 nm were primarily investigated. In the present study, grain growth behavior is examined in grain size regime below 10 nm in nanocrystalline Ni-W alloys with 8 to 15 at% W at 773K. In this size regime, normal grain growth behavior was observed in these alloys, evidenced from (a) parabolic kinetics in plots of square of average grain size against annealing time, and (b) statistical self-similarity on the scaled size distribution in a log-normal plot.

## 1. Introduction

Studies on grain growth in nanocrystalline(nc) materials have revealed interesting problems, particularly associated with the thermodynamics and kinetics of grain boundaries. One of the well-known issues with nanocrystalline metals is their poor grain size stability which can be understood from eqn. 1:

$$dG = \gamma dA \tag{1}$$

where $dG$ is the change in Gibbs free energy, $\gamma$ is the grain boundary energy and $dA$ is change in grain boundary (GB) area. In the nanocrystalline regime, where the grain is size less than 100 nm, the volume fraction of the grain boundary region increases, as demonstrated by Palumbo et al. [1], leading to huge driving forces for grain growth.

Grain growth is broadly divided into normal and abnormal grain growth. The microstructure grows uniformly during normal grain growth, also considered to be a continuous process, characterized by (a) a parabolic growth law (i.e., the square of the average grain size increases linearly with time) and (b) the scaled grain size distribution (in which grain size is normalized by its mean value) remains invariant with time; therefore, this process is also termed as self-similar grain growth [2]. On the contrary, abnormal grain growth is considered to be a discontinuous process where the microstructure acquires more than one characteristic grain size.

The conditions implicit in the classical theory of normal grain growth are: (a) GB migration follows the equation: $v = M\kappa\gamma$ where the migration velocity $v$ is proportional to the driving force = $\kappa\gamma$, with $\kappa$ as the mean curvature and $\gamma$ is the grain boundary energy and $M$ is the GB mobility, and (b) statistical self-similarity, wherein the grain growth occurs in a geometrically similar way in a statistical sense, obtained from scaled grain size distribution during grain growth [3,4]. It also implies that all the grain boundaries are essentially the same, irrespective of grain size or misorientation.

Early studies on nanocrystalline FCC pure metals have reported abnormal grain growth at lower homologous temperatures [5-9], where some grains grow with significantly different driving forces and velocity and, therefore, the material acquires more than one characteristic grain size. For example, Klement et al. [5] reported abnormal grain growth in nanocrystalline Ni where the as-deposited grain size of 10 and 20 nm increased at a temperature as low as 443 K (0.3 homologous temperature) to 40 and 100 nm. Gertsman and Birringer [6] suggested that inhomogeneity of grain boundary structure and nonuniform interface segregation may result in the abnormal grain growth observed at room temperature in nanocrystalline copper. Subsequent reports [5-9] attributed an abnormal or anomalous grain growth in nc-Ni to the segregation of sulfur at the grain boundary or formation of sulfur-rich second phase $Ni_3S_2$.

The addition of solutes in Ni-based nanocrystalline alloys have improved their thermal stability. In 1993, Weissmüller [10] first showed that the grain boundaries can be stabilized by solute segregation in the nanocrystalline regime. Subsequently, the alloying effects in nanostructures and their thermal stability have been studied widely in many Ni-based alloys [11-21].

Generally, grain growth among Ni-based alloys has been observed to occur in two stages: (a) a lower growth rate at low homologous temperatures and (b) increased growth rate after a transition temperature. The transition temperature increases with the addition of solute and kind of solute. For example, the transition temperature is 573 K for nanocrystalline Ni and increases to 773 K with addition of W. In several reports for nanocrystalline Ni-Fe and Ni-W alloys [11-24], a lower growth rate in the first stage is attributed to the relaxation of grain boundaries accompanied by the release of micro-strain as measured by XRD and a shallow peak in a DSC curve. The second stage is accompanied by either normal or abnormal/accelerated grain growth depending on the solute. Table 1 lists information on the transition temperature and the type of grain growth observed in the second stage. Though the addition of solutes such as Fe, Co, and Mn showed improved thermal stability, these alloys exhibited limited or complete abnormal grain growth [11-18]. While nanocrystalline Ni-Mo and Ni-W alloys [19-24] show superior thermal stability (0.4 homologous temperature) and abnormal grain growth has not been reported among these alloys. The present study focuses on nanocrystalline Ni-W alloys.

Table 1: Grain growth behaviour of various nanocrystalline Ni binary alloys

| Alloy (at%) | $d_o$ (nm) | T, time (K, minutes) | $d_f$ (nm) | Type of grain growth | Ref. |
|---|---|---|---|---|---|
| Ni-15Fe | 15 | 523, 90 | 24 & 90 | Bimodal grain size distribution | [11-13] |
| Ni-15Fe | 15 | 723, 90 | 190 | Limited abnormal growth | [11-13] |
| Ni-37Co | 12 | 573, 120 | 400 | Abnormal grain growth | [14-16] |
| Ni-52Co | 12 | 573, 120 | 400 | Abnormal grain growth | [14-16] |
| Ni-1.2Mn | 18 | 673, 60 | 90 | Higher thermal stability and NGG | [17, 18] |
| Ni-13Mo | 6 | 823, 60 | 25 | Higher thermal stability and normal grain growth | [19, 20] |
| Ni-17Mo | 5 | 823, 60 | 15 | Higher thermal stability and normal grain growth | [19, 20] |

Detor and Schuh [21] studied grain boundary segregation and chemical ordering during grain growth in nanocrystalline Ni-W alloys across a wide range of grain sizes, from 3 to 70 nm, to evaluate grain size stability in different temperature regimes. They showed that grain size remains stable at temperatures up to 773 K over 24 hours and normal grain growth occurs at higher temperatures. Though one of the as-deposited grain size was 3 nm, variation in grain size distribution upon annealing was not probed for grain sizes below < 10 nm. Matsui et al. [22] reported grain growth behavior in Ni-W alloys for grain size (d>27 nm), annealed between 573 to 773 K. They reported the absence of second phase and oxide particle and precipitates as reported by Marvel et al [23] in Ni-23 at% W, annealed up to 973 K. Schuler et al. [24] found a new high-temperature stability region in Ni-5at% W where they reported a lack of dramatic grain growth after 1373 K and attributed it to the formation of amorphous intergranular films (AIF). There is a difference in the temperature at which grain growth starts and second phase appears, among various nanocrystalline Ni-W alloys being reported. These differences are within 100 K and could be due to changes in deposition parameters and solute content. Significant grain growth occurs close to 773K in nanocrystalline Ni-W alloys and second phases start appearing at or above 873 K depending on the W content.

Nanocrystalline Ni-W alloys exhibit thermal stability up to a homologous temperature of 0.4 (773 K). Most of the reports focus on thermal stability and the grain growth behavior has only been investigated at higher temperatures in the grain size regime greater than 15-20 nm. The aim of the present study was to systematically study grain growth behavior in nanocrystalline Ni-W alloys for grain sizes below 10 nm.

## 2. Materials and methods

Three Ni-W alloys with different W content were produced via electrodeposition; details of the electrodeposition parameters are given elsewhere [25]. The Ni and W content were analysed using wavelength dispersive spectroscopy (WDS) on an Electron Probe Micro Analyser (EPMA), from JEOL (JXA-8530F). Alloys were checked for C, O and S impurities from two kinds of measurements: (i) CHNS/O micro analyzer and (ii) Atom probe tomography (APT).

The chemical analysis of the three as-deposited alloys revealed 3.3, 8.4 and 15.2 at% W, and these are henceforth referred to as Ni-3 W, Ni-8W and Ni-15W, respectively. Sulfur was absent in these alloys; further, carbon and oxygen content were found to be below 0.02 at %.

XRD measurements were performed on Ni-W alloys using Cu K$_\alpha$ radiation at 40 kV and 30 mA, with a PANalytical X'Pert Pro instrument. The Scherrer method [26] was used to calculate the grain size, following the equation, $B(2\theta) = \frac{K\lambda}{Lcos\theta}$ where $B$ is the peak width i.e. FWHM of 111 XRD peak, $K$ = 0.9, $2\theta$ is the Bragg angle, $\lambda$ is the wavelength of Cu K$_\alpha$ and $L$ is the crystallite size. Transmission electron microscopy (TEM) investigations were performed with a 300 kV FEI TITAN (without aberration correction) on samples prepared by a GATAN Precision Ion Polishing System (PIPS) [25].

## 3. Results

The three alloys were annealed for 60 minutes at temperatures ranging from 373 to 1073 K. No grain growth was observed below 773 K; further, at and above 873 K, second phase precipitation was observed in Ni-15W alloy. In the alloy Ni-3W, a two-fold increase in GB segregation was observed during annealing in an earlier study in our group, while GB segregation remained unchanged in Ni-8W and Ni-15W [25]. Therefore, the grain growth studies were restricted to Ni-8W and Ni-15W at 773 K; since, the primary interest is to access grain sizes smaller than 10 nm, these alloys were given short annealing treatment for 20, 60 and 120 minutes.

The grain sizes in the as-deposited and annealed samples of Ni-8W and Ni-15W, determined using the Scherrer method, are listed in Table 2 (the XRD patterns are shown in Fig S1 and S2 of the supplementary file). The as-deposited grain sizes of Ni-8W and Ni-15W were 8.4 and 4.9 nm, respectively.

### 3.1 Grain growth

The square of the average grain size is plotted against time in Fig. 1 for the as-deposited and annealed Ni-8W and Ni-15W samples. A linear behaviour implying a parabolic growth law (predicted for normal grain growth) is observed in Ni-15W until 120 minutes. However, a linear behaviour is observed only until 60 minutes in Ni-8W beyond which grain growth is hastened; while the grain size increased from 8.4 nm to 10 nm during the first 60 minutes, it increased further to 22 nm in the next 60 minutes, leading to the appearance of a kink in Fig. 1 for this alloy. We discuss this further in the discussion section.

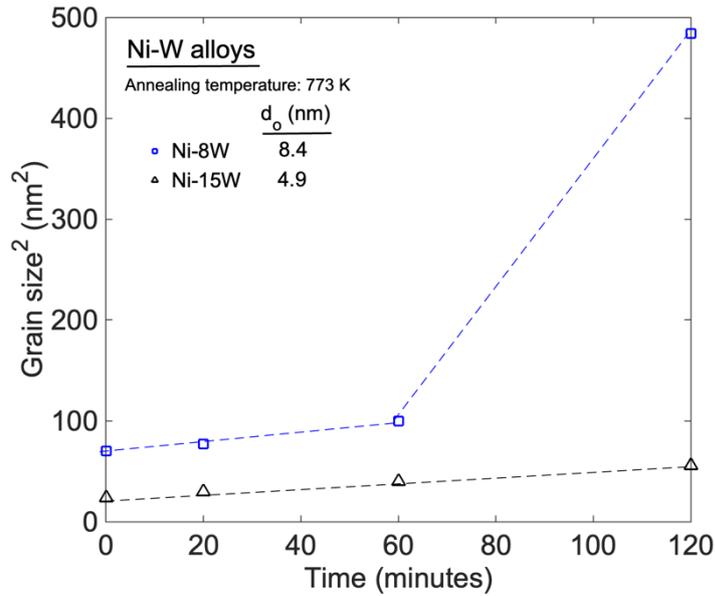

Fig. 1: Variation of square of grain size upon annealing at 773 K with time: 20, 60 and 120 minutes

3.2 Grain size distribution

To examine the details of this normal grain growth regime (which is observed during 60 minutes of annealing), an investigation of grain size distribution was undertaken using TEM observations. Dark field images of the as-deposited and annealed (773 K, 60 minutes) Ni-8W and Ni-15W alloys are shown in Fig. 2 and Fig. 3.

Twin boundaries were observed in many grains in these alloys; their presence was noted in dark field images (in Fig. 2 and 3) and also confirmed using bright field and high resolution images (in Fig. S3 in the supplementary file). Dark field TEM images were taken after placing the selected area diffraction (SAD) aperture on the double spots in the diffraction pattern shown in the inset of Fig 2 (a); twin boundaries in these images are marked using turquoise-colored circles.

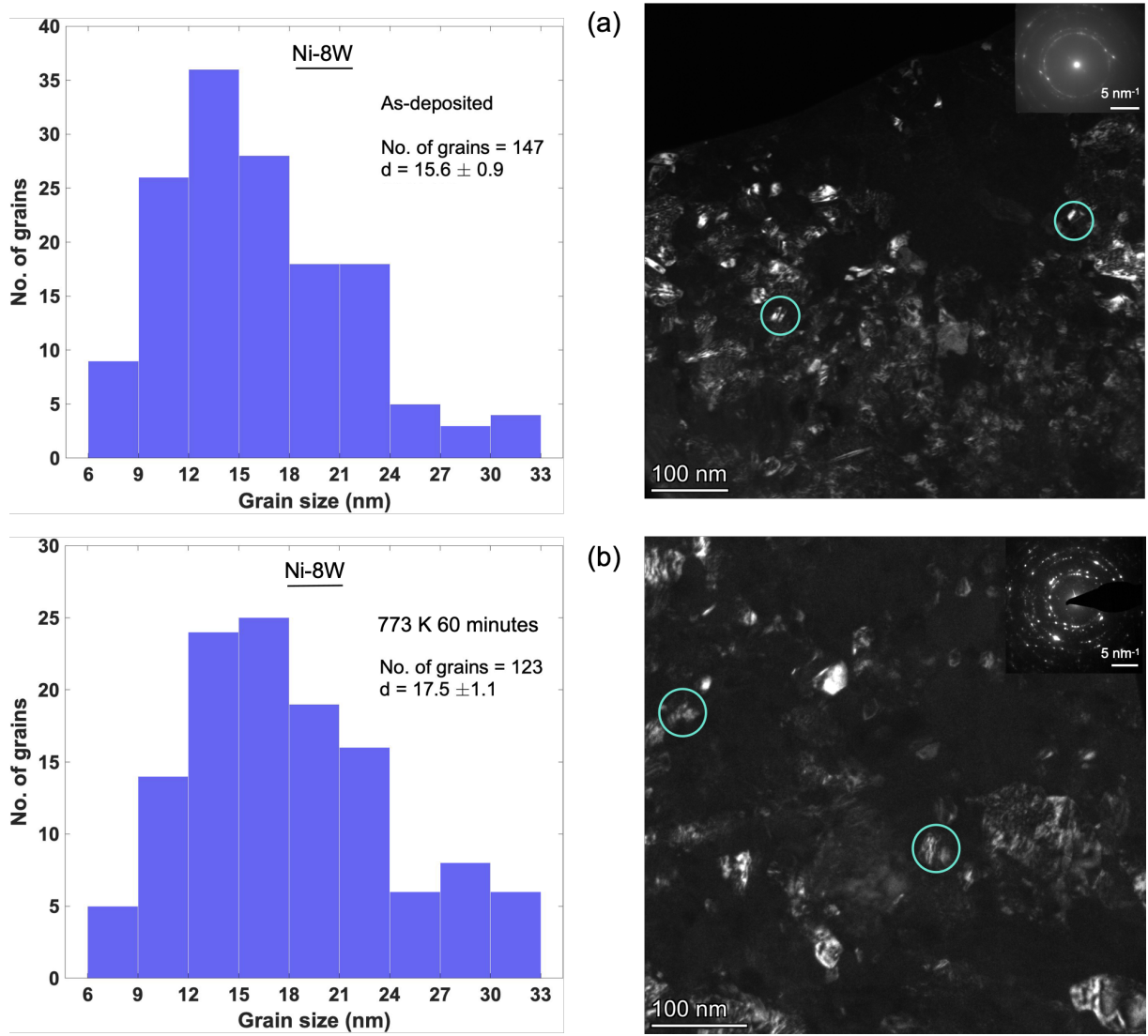

Fig. 2: Histograms of grain size and their frequency of Ni-8W obtained from ImageJ analysis of dark field images of (a) as-deposited (b) annealed at 773 K for 60 minutes. The inset in histogram contains information on number of grains and TEM grain size excluding twin boundaries. The marked regions within turquoise-colored circles highlight the presence of twin boundaries.

The sizes of individual grains in the dark field images (with and without twin boundaries) were measured using ImageJ. The TEM and Scherrer grain size of Ni-8W and Ni-15W are noted in Table 2. The Scherrer grain size for these alloys are close to the grain size calculated from TEM including TBs; this is expected since the XRD measurements account for coherently scattered domains [27].

Due to their low mobility, TBs do not contribute to grain growth; in other words, grain growth depends mainly on the migration of GBs. Therefore, we use grain sizes obtained from the dark field images without considering TBs for plotting the size distributions in the histograms in Fig. 2 and 3.

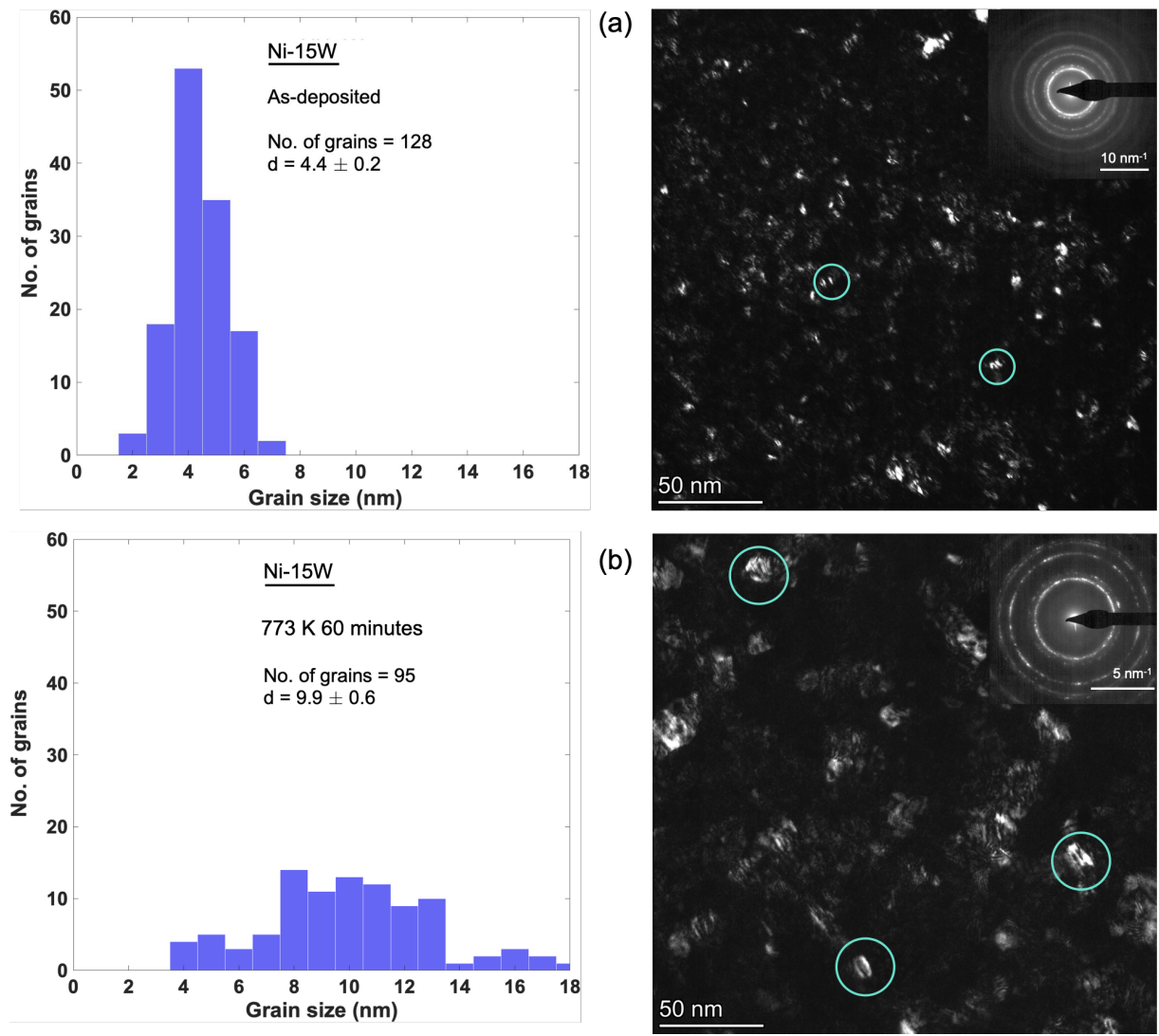

Fig. 3: Histograms of grain size and their frequency of Ni-15W obtained from ImageJ analysis of dark field images of (a) as-deposited (b) annealed 773 K for 60 minutes. The inset in histogram contains information on number of grains and TEM grain size excluding twin boundaries. The marked regions within turquoise-colored circles highlight the presence of twin boundaries.

In Fig. 2, for Ni-8W, the peak in the histogram shifts to the right accompanied by an increase in the number of large grains, leading to an increase in the average grain size from 15.6 to 17.5 nm. In the case of Ni-15W, the size distribution shows a substantial broadening along with a comparatively larger rightward shift in the peak leading to a doubling of average grain size (4.4 to 9.9 nm).

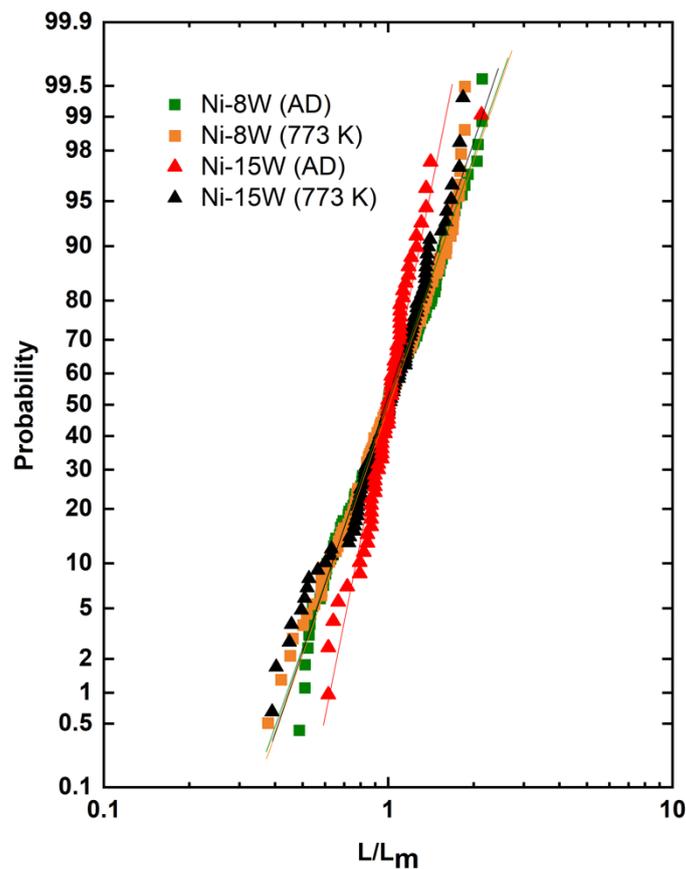

Fig. 4: Grain size distribution in Ni-8W and Ni-15W alloys upon annealing [The markers for Ni-W alloys have been used with a different color than the rest of the images for visual clarity. In the legend, AD refers to as-deposited samples and 773 K refers to annealed samples at 773 K for 60 minutes]

In order to test for the existence of statistical self-similarity (which accompanies normal grain growth), the grain size distribution is re-plotted in Fig. 4 in a log-normal plot in which the cumulative probability is plotted against the logarithm of normalised grain size (i.e., individual grain size divided by the average grain size). Since the grain sizes are normalized, an overlap of two scaled size distributions would imply statistical self-similarity. Data falling on a straight line in such plots would imply that the grain size distribution is log-normal. TEM grain size as noted in Table 2 was used for the above analysis which is the grain size obtained from dark field images excluding twin boundaries.

Table 2: TEM and XRD grain sizes of Ni-8W and Ni-15W. TEM grain sizes (under the column entitled TEM) and including TBs (designated TB-TEM) noted along with $S_n$ i.e. the standard deviation.

| Alloy | XRD d (nm) | TEM d (nm) | TB-TEM d (nm) | $S_n$ (nm) | No. of grains |
|---|---|---|---|---|---|
| Ni-8W (AD) | 8.4 | 15.6 ± 0.9 | 9.0 ± 0.9 | 5 | 147 |
| Ni-8W (773 K 60 minutes) | 10 | 17.5 ± 1.1 | 11.8 ± 1.2 | 6 | 123 |
| Ni-15W (AD) | 4.9 | 4.4 ± 0.2 | 3.9 ± 0.2 | 1 | 128 |
| Ni-15W (773 K 60 minutes) | 6.3 | 9.9 ± 0.6 | 5.9 ± 0.6 | 3 | 95 |

## 4. Discussion

Given the presence of substantial driving forces at grain sizes below 100 nm, normal grain growth is not expected in this regime. Nanocrystalline pure elements and a few alloys, mentioned in Table 1, show abnormal grain growth in different grain size and temperature regime. While nanocrystalline Ni-W and Ni-Mo have shown higher thermal stability, details of grain growth behavior remain unexplored. The present study shows a normal grain growth behavior along with statistical self-similarity for Ni-W alloys at grain sizes below 10 nm.

### 4.1 Parabolic growth law:

A report by Detor and Schuh [21] investigated grain sizes below 10 nm among nanocrystalline Ni-W alloys - 3 nm (21 at. %W), 20 nm (13 at. %W), and 70 nm (6 at. %W). These alloys were annealed in the range 423 to 1173 K for a duration of 0.5 to 72 hours. The 3 nm alloy was annealed for 24 hours from 573 to 1173 K; the grain size gradually changed to 11 nm at 723 K and a quickened growth rate from 14 to 80 nm from 873 to 1073 K. Though they considered as-deposited grain size of 3 nm, they did not investigate grain growth behavior by analyzing the time-dependence of growth rate and change of grain size distribution.

In the present study, Ni-8W and Ni-15W were annealed at 773 K for 20, 60 and 120 minutes. Ni-15W alloy follows a linear behavior up to 120 minutes, on a plot of the square of the average grain size against time, as shown in Fig. 1, consistent with normal grain growth. A kink is observed for the Ni-8W alloy beyond 60 minutes. A similar kink was observed in nanocrystalline Ni by Natter et.al [8] at 503 K after t > 200 minutes. Under the assumption of monomodal grain size distribution, the observed kink is reported to occur upon the broadening of the grain size distribution. In the present study, both the Ni-W alloys have monomodal grain size distribution, as shown in Fig. 2 and 3. It follows from the above observation that the observed kink in Ni-8W alloy is possibly due to the broadening of grain size distribution after t > 60 minutes.

### 4.2 Statistical self-similarity:

Two samples from each of these alloys were chosen for analyzing the grain size distribution variation upon annealing – one as-deposited and the other with 60 minutes of annealing.

The standard deviation, $S_n$ values range from 1-6 nm in the present study in contrast to pure nanocrystalline Ni and Ni-Fe, Co, Mn alloys where the standard deviation is higher between 10 to 25 nm. It is expected for the grain size distribution to be larger during abnormal grain growth [2].

All the alloys fall on a single line, in Fig. 4, with a slight deviation observed in the as-deposited Ni-15W. To understand this deviation, we looked at the normalised standard deviation (standard deviation/mean grain size) for all the alloys, as mentioned in Table 2. The values of normalised standard deviation is close to 0.3 for Ni-8W alloys and annealed Ni-15W while it is equal to 0.2 for as-deposited Ni-15W. The possible reason for the observed difference in normalised standard deviation value could be the absence of information on the smaller grain sizes (< 3 nm) in the distribution, due to the limit of resolution in TEM. Absence of information on lower grain sizes may have led to a smaller $S_n$ value of 1 for Ni-15W alloys as compared to Ni-8W where $S_n$ = 5,6.

The as-deposited Ni-15W data could fall on the single line like rest of these alloys in Fig. 4, if these grain sizes were to be accounted for, implying a self-similar grain growth in nanocrystalline Ni-W alloys. Since, the Ni-15W alloy followed the parabolic grain growth, as shown in Fig. 1, which assumes self-similarity [2], therefore it could be argued that the observed deviation in the as-deposited Ni-15W, be ignored.

## 5. Summary

In summary, the probability vs log (normalized grain size) plot of as-deposited and annealed Ni-8W and Ni-15W alloys revealed that the microstructures had a log-normal grain size distribution and statistical self-similarity. The parabolic grain growth as shown in Fig. 1, in this regime suggests isotropic grain boundary properties [2]. Following these observations and from our earlier work, where W was shown to uniformly segregate at the grain boundaries [25], it is suggested that the grain boundary mobility and energy are uniform among these alloys.

## Acknowledgement:

The authors would like to express their gratitude to Prof. A H. Chokshi and Prof. T. A. Abinandanan for several discussions and critical review of this work.

**Data availability:**

The raw/processed data required to reproduce the above findings cannot be shared at this time as the data also forms part of an ongoing study and data will be made available on request.